\documentclass[twocolumn,floatfix,showpacs]{revtex4}
\usepackage{graphicx}
\usepackage{amsmath}
\usepackage{amssymb}
\usepackage{bm}

\begin{document}

\title{Domain wall in a chiral \textit{p}-wave superconductor: a pathway for electrical current}
\author{I. Serban}
\affiliation{Instituut-Lorentz, Universiteit Leiden, P.O. Box 9506, 2300 RA Leiden, The Netherlands}
\author{B. B\'{e}ri}
\affiliation{Instituut-Lorentz, Universiteit Leiden, P.O. Box 9506, 2300 RA Leiden, The Netherlands}
\author{A. R. Akhmerov}
\affiliation{Instituut-Lorentz, Universiteit Leiden, P.O. Box 9506, 2300 RA Leiden, The Netherlands}
\author{C. W. J. Beenakker}
\affiliation{Instituut-Lorentz, Universiteit Leiden, P.O. Box 9506, 2300 RA Leiden, The Netherlands}
\date{December 2009}
\begin{abstract}
Superconductors with $p_{x}\pm ip_{y}$ pairing symmetry are characterized by chiral edge states, but these are difficult to detect in equilibrium since the resulting magnetic field is screened by the Meissner effect. Nonequilibrium detection is hindered by the fact that the edge excitations are unpaired Majorana fermions, which cannot transport charge near the Fermi level. Here we show that the boundary between $p_{x}+ip_{y}$ and $p_{x}-ip_{y}$ domains forms a one-way channel for electrical charge. We derive a product rule for the domain wall conductance, which allows to cancel the effect of a tunnel barrier between metal electrodes and superconductor and provides a unique signature of topological superconductors in the chiral \textit{p}-wave symmetry class.
\end{abstract}
\maketitle

Chiral edge states are gapless excitations at the boundary of a two-dimensional system that can propagate in only a single direction. They appear prominently in the quantum Hall effect \cite{Hal82,But88}: The absence of backscattering in a chiral edge state explains the robustness of the quantization of the Hall conductance against disorder. Analogous phenomena in a superconductor with broken time reversal symmetry are known as the spin quantum Hall effect \cite{Vol89,Sen99,Rea00} and the thermal quantum Hall effect \cite{Sen00,Vis01}, in reference to the transport of spin and heat along chiral edge states. 

Unlike the original (electrical) quantum Hall effect, both these superconducting analogues have eluded observation, which is understandable since it is so much more difficult to measure spin and heat transport than electrical transport. Proposals to detect chiral edge states in a superconductor through their equilibrium magnetization are hindered by screening currents in the bulk, which cancel the magnetic field (Meissner effect) \cite{Mat99,Kwo03,Kir07,Kal09}.

\begin{figure}[tb]
\centerline{\includegraphics[width=0.9\linewidth]{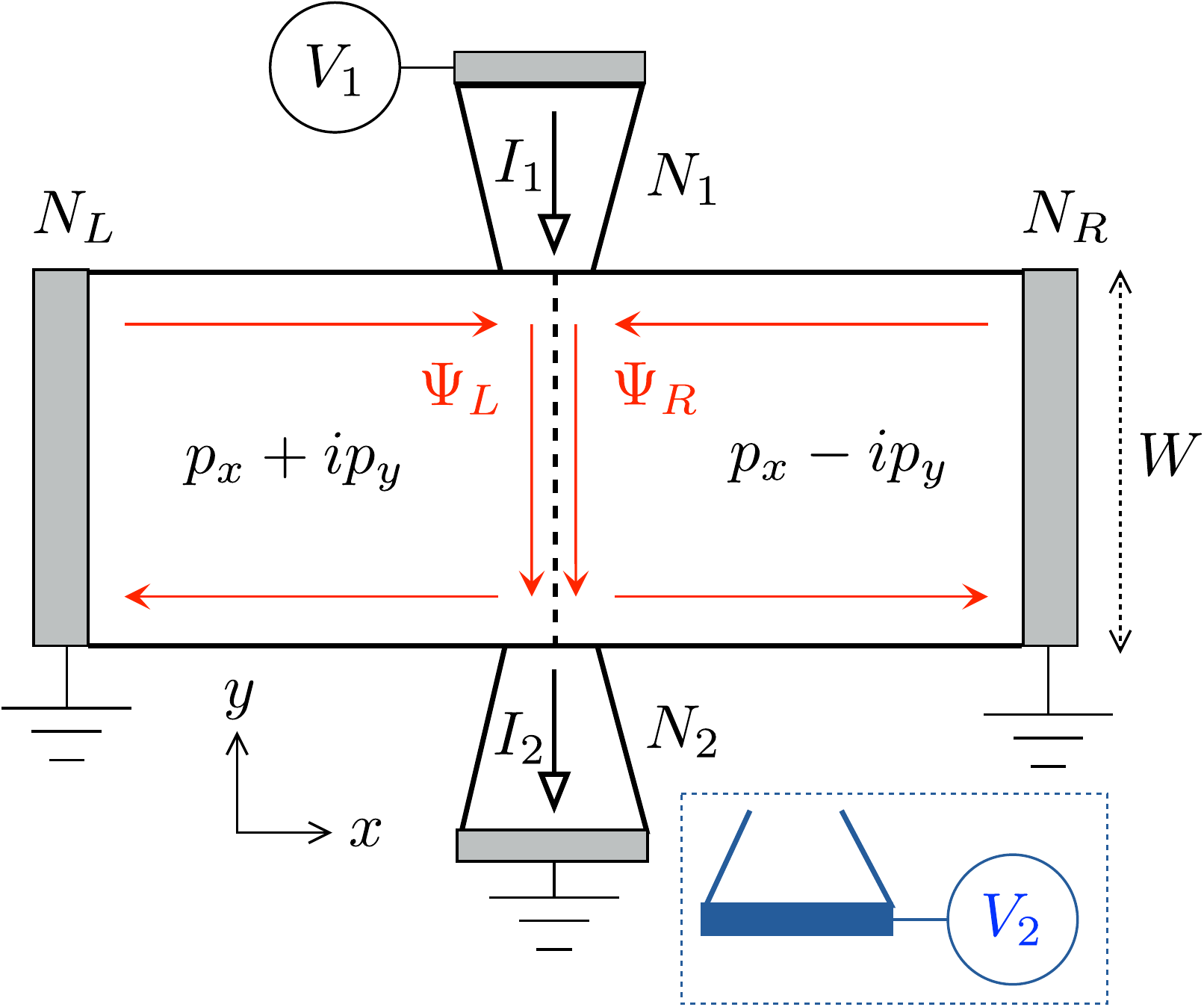}}
\caption{\label{fig_schematic}
Superconducting strip divided by a domain wall (dashed line, length $W$) into domains with $p_{x}\pm ip_{y}$ symmetry. The edge states $\Psi_{L},\Psi_{R}$ of opposite chirality in the two domains are indicated by red arrows. These unpaired Majorana modes can carry heat current between contacts $N_{L}$ and $N_{R}$, but no electrical current. A normal metal electrode $N_{1}$ at voltage $V_{1}$ injects charge into the domain wall, which is detected as an electrical current $I_{2}$ at the other end $N_{2}$. In an alternative measurement configuration (indicated in blue), contact $N_{2}$ measures a voltage $V_{2}$ without drawing a current.
}
\end{figure}

Here we show that the boundary between domains of opposite chirality ($p_{x}\pm ip_{y}$) in a chiral \textit{p}-wave superconductor forms a one-way channel for electrical charge, in much the same way as edge states in the quantum Hall effect. This is not an immediate consequence of chirality: Since the charge of excitations in a superconductor is only conserved modulo the Cooper pair charge of $2e$, the absence of backscattering in a superconducting chiral edge state does not imply conservation of the electrical current. Indeed, one chiral edge state within a single domain has zero conductance due to electron-hole symmetry. We calculate the conductance of the domain wall, measured between a pair of metal contacts at the two ends (see Fig.\ \ref{fig_schematic}), and find that it is nonzero, regardless of the separation of the contacts.

Our analysis is generally applicable to so-called class-D topological superconductors \cite{Alt97,Sch08}, characterized by the presence of electron-hole symmetry and the absence of both time-reversal and spin-rotation symmetry. It can be applied to the various realizations of chiral \textit{p}-wave superconductors proposed in the literature (strontium ruthenate \cite{Kal09}, superfluids of fermionic cold atoms \cite{Tew07,Sat09}, and ferromagnet-superconductor heterostructures \cite{Sau09,Lee09}).

We start from the  Bogoliubov-De Gennes equation,
\begin{equation}
\begin{pmatrix}
H_{0}-E_{F}&\Delta\\
\Delta^{\dagger}&-H_{0}^{\ast}+E_{F}
\end{pmatrix}
\begin{pmatrix}
u\\
v
\end{pmatrix}=E
\begin{pmatrix}
u\\
v
\end{pmatrix},\label{Hdef}
\end{equation}
for coupled electron and hole excitations $u(\bm{r}),v(\bm{r})$ at energy $E$ above the Fermi level $E_{F}$. The single-particle Hamiltonian is $H_{0}=(\bm{p}+e\bm{A})^{2}/2m+U$, with $\bm{p}=-i\hbar\partial/\partial\bm{r}$ the momentum, $\bm{A}(\bm{r})$ the vector potential, and $U(\bm{r})$ the electrostatic potential. The dynamics is two-dimensional, so $\bm{r}=(x,y)$, $\bm{p}=(p_{x},p_{y})$. The pair potential $\Delta$ has the spin-polarized-triplet \textit{p}-wave form \cite{Ste04}
\begin{equation}
\Delta=(2p_{F})^{-1}(\bm{\eta}\cdot\bm{p}+\bm{p}\cdot\bm{\eta}),\label{Deltadef}
\end{equation}
in terms of a two-component order parameter $\bm{\eta}=(\eta_{x},\eta_{y})$. The two chiralities $p_{x}\pm ip_{y}$ correspond to $\bm{\eta}_{\pm}=\Delta_{0}e^{i\phi}(1,\pm i)$, with $\Delta_{0}$ the excitation gap and $\phi$ the superconducting phase. Since $\Delta^{\dagger}=-\Delta^{\ast}$, a solution $(u,v)$ of Eq.\ \eqref{Hdef} at energy $E$ is related to another solution $(v^{\ast},u^{\ast})$ at energy $-E$ (electron-hole symmetry). A domain wall along $x=0$, with a phase difference $\phi$ between the domains, has order parameter \cite{Sig99,Bou09}
\begin{subequations}
\label{etadef}
\begin{align}
&\eta_{x}(x)=\Delta_{0}[e^{-i\phi/2}\cos\chi(x)+e^{i\phi/2}\sin\chi(x)],\label{etaxdef}\\
&\eta_{y}(x)=i\Delta_{0}[e^{-i\phi/2}\cos\chi(x)-e^{i\phi/2}\sin\chi(x)],\label{etaydef}
\end{align}
\end{subequations}
The function $\chi(x)$ increases from $0$ to $\pi/2$ over a coherence length $\xi_{0}=\hbar v_{F}/\Delta_{0}$ around $x=0$.

At energies $E$ below $\Delta_{0}$ the excitations are nondegenerate chiral edge states $\Psi_{L}$ and $\Psi_{R}$ circulating in opposite directions in the two domains \cite{Mat99,Bar01,Kwo04,Sto04}. (See Fig.\ \ref{fig_schematic}.) At the domain wall the two states mix, so that an excitation entering the domain wall in the state $\Psi_{L}^{\rm in}$ or $\Psi_{R}^{\rm in}$ can exit in either of the two states $\Psi_{L}^{\rm out}$ and $\Psi_{R}^{\rm out}$. We first analyze this edge state scattering problem between contacts $N_{L}$ and $N_{R}$, and then introduce the contacts $N_{1}$ and $N_{2}$ to the domain wall. 

The edge state excitations have fermionic annihilation operators $\bm{\gamma}(E)=\bigl(\gamma_{L}(E),\gamma_{R}(E)\bigr)$, which satisfy the electron-hole symmetry relation
\begin{equation}
\bm{\gamma}(E)=\bm{\gamma}^{\dagger}(-E).\label{ehsymmetry}
\end{equation}
At zero energy one has $\bm{\gamma}=\bm{\gamma}^{\dagger}$, so these are Majorana fermions \cite{Ste04}. The unitary scattering matrix $S(E)$ relates incoming and outgoing operators, $\bm{\gamma}^{\rm out}(E)=S(E)\bm{\gamma}^{\rm in}(E)$. Electron-hole symmetry for both $\bm{\gamma}^{\rm in}$ and $\bm{\gamma}^{\rm out}$ requires $S(E)\bm{\gamma}^{\rm in}(E)=\bm{\gamma}^{\rm in}(E)S^{\dagger}(-E)$, hence $S(E)=S^{\ast}(-E)$. The zero-energy scattering matrix $S(0)\equiv S_{dw}$ of the domain wall is therefore a real unitary, or orthogonal, matrix. We may parametrize it by
\begin{equation}
S_{dw}=\begin{pmatrix}
\cos\psi&\sin\psi\\
(-1)^{p+1}\sin\psi&(-1)^{p}\cos\psi
\end{pmatrix}=\sigma_{z}^{p}e^{i\psi\sigma_{y}},\label{Salpha}
\end{equation}
in terms of a mixing angle $\psi$ and a parity index $p\in\{0,1\}$.

The mixing angle $\psi=k_{y}W$ is determined by the phase accumulated by the pair of chiral Majorana modes, as they propagate with wave number $\pm k_{y}$ along the domain wall of length $W$. The dispersion relation $E(k_{y})$ of the Majorana modes was calculated in Ref.\ \cite{Kwo04}, for a step function order parameter at $x=0$, including also the effect of a tunnel barrier $U=U_{0}\delta(x)$ (tunnel probability $D$, zero magnetic field). By equating $E(k_{y})=0$ and solving for $k_{y}$ we obtain the mixing angle
\begin{equation}
\psi=k_F W\sqrt{D}\cos(\phi/2).\label{alpharesult}
\end{equation}
The mixing angle can in principle be measured through thermal transport between contacts $N_{L}$ and $N_{R}$, since the heat current through the domain wall is $\propto\sin^{2}\psi$. In what follows we consider instead a purely electrical measurement of transport along the domain wall, that (as we shall see) is independent of the degree of mixing of the Majorana modes. 

The measurement that we propose consists of the injection of electrons from contact $N_{1}$ at voltage $V_{1}$ (relative to the superconductor) and the detection at contact $N_{2}$. We consider two detection schemes: In the first scheme contact $N_{2}$ is kept at the same potential as the superconductor and measures a current $I_{2}$, leading to the nonlocal conductance $G_{12}=I_{2}/V_{1}$. In the second scheme contact $N_{2}$ is a voltage probe drawing no net current and measuring a voltage $V_{2}$. The ratio $R_{12}=V_{2}/I_{1}$, with $I_{1}$ the current entering the superconductor through contact $N_{1}$, is the nonlocal resistance. The two nonlocal quantities are related by $R_{12}=G_{12}/G_{1}G_{2}$, with $G_{i}=|I_{i}/V_{i}|$ the contact conductance of electrode $N_{i}$ (measured with the other contact grounded). 

We take the zero-temperature and zero-voltage limit, so that we can use the zero-energy scattering matrix to calculate the various conductances. The scattering problem at contact $N_{1}$ involves, in addition to the Majorana operators $\bm{\gamma}=(\gamma_{L},\gamma_{R})$, the electron and hole annihilation operators $a_{n}$ and $b_{n}$ in mode $n=1,2,\ldots N$. These are related by $b_{n}(E)=a_{n}^{\dagger}(-E)$. The even and odd combinations $\gamma^{\pm}_{n}$, defined by
\begin{equation}
\begin{pmatrix}
\gamma^{+}_{n}\\
\gamma^{-}_{n}
\end{pmatrix}=u\begin{pmatrix}
a_{n}\\
b_{n}
\end{pmatrix},\;\;
u=\sqrt{\frac{1}{2}}\begin{pmatrix}
1&1\\
-i&i
\end{pmatrix},\label{gammadef}
\end{equation}
satisfy the same electron-hole symmetry relation \eqref{ehsymmetry} as $\gamma_{L},\gamma_{R}$, and therefore represent Majorana fermions at $E=0$. We denote $\bm{\gamma}_{n}=(\gamma^{+}_{n},\gamma^{-}_{n})$ and collect these operators in the vector $\bm{\Gamma}=(\bm{\gamma}_{1},\bm{\gamma}_{2},\ldots \bm{\gamma}_{N})$. The scattering matrix $S_{1}$ of contact $N_{1}$ relates incoming and outgoing operators,
\begin{align}
\begin{pmatrix}
\bm{\gamma}\\
\bm{\Gamma}
\end{pmatrix}_{\rm out}
=S_{1}
\begin{pmatrix}
\bm{\gamma}\\
\bm{\Gamma}
\end{pmatrix}_{\rm in},\;\;
S_{1}=
\begin{pmatrix}
r_{1}&t_{1}\\
t'_{1}&r'_{1}
\end{pmatrix}.\label{S1def}
\end{align}
Electron-hole symmetry implies that $S_{1}$ is $(2N+2)\times(2N+2)$ orthogonal matrix at zero energy. Similarly, the zero-energy scattering matrix $S_{2}$ of contact $N_{2}$ is a $(2N'+2)\times (2N'+2)$ orthogonal matrix. (The number of modes is $N,N'$ in contacts $N_{1},N_{2}$ respectively.)

The $2N'\times 2N$ transmission matrix 
\begin{equation}
t_{21}=t'_{2}S_{dw}t_{1}=t'_{2}\sigma_{z}^{p}e^{i\psi\sigma_{y}}t_{1}\label{t21def}
\end{equation}
from contact $N_{1}$ to $N_{2}$ is the product of the $2\times 2N$ submatrix $t_{1}$ of $S_{1}$ (transmission from $N_{1}$ to the domain wall), the $2\times 2$ scattering matrix $S_{dw}$ (transmission along the domain wall), and the $2N'\times 2$ submatrix $t'_{2}$ of $S_{2}$ (transmission from the domain wall to $N_{2}$). 

The total transmission probability $T_{ee}$, summed over all modes, of an electron at contact $N_{1}$ to an electron at contact $N_{2}$ is given by
\begin{align}
T_{ee}&=\tfrac{1}{4}{\rm Tr}\,{\cal U}^{\dagger}t_{21}^{\dagger}{\cal U}(1+\Sigma_{z}){\cal U}^{\dagger}t_{21}{\cal U}(1+\Sigma_{z})\\
&=\tfrac{1}{4}{\rm Tr}\,t_{21}^{\dagger}(1-\Sigma_{y})t_{21}(1-\Sigma_{y})
,\label{Teedef}
\end{align}
where we have defined the direct sums ${\cal U}=u\oplus u\cdots\oplus u$, $\Sigma_{i}=\sigma_{i}\oplus\sigma_{i}\cdots\oplus\sigma_{i}$ and we have used that $u\sigma_{z}u^{\dagger}=-\sigma_{y}$. Similarly, the total electron-to-hole transmission probability $T_{he}$ reads
\begin{equation}
T_{he}=\tfrac{1}{4}{\rm Tr}\,t_{21}^{\dagger}(1+\Sigma_{y})t_{21}(1-\Sigma_{y}).\label{Thedef}
\end{equation}

Since $I_{2}=(e^{2}/h)V_{1}(T_{ee}-T_{he})$, the nonlocal conductance takes the form
\begin{equation}
G_{12}=(e^{2}/h)\tfrac{1}{2}{\rm Tr}\,t_{21}^{T}\Sigma_{y}t_{21}\Sigma_{y}.\label{Gdef}
\end{equation}
We have used that $t_{21}^{\dagger}=t_{21}^{T}$ and ${\rm Tr}\,t_{21}^{T}\Sigma_{y}t_{21}=0$ (being the trace of an antisymmetric matrix). The nonlocal resistance can be written in a similar form upon division by the contact conductances,
\begin{equation}
R_{12}=\frac{G_{12}}{G_{1}G_{2}},\;\;G_{i}=(e^{2}/h)\tfrac{1}{2}{\rm Tr}\,(1-\Sigma_{y}{r'}_{i}^{T}\Sigma_{y}r'_{i}).\label{Gidef}
\end{equation}
We will henceforth set $e^{2}/h$ to unity in most equations.

Substitution of Eq.\ \eqref{t21def} into Eq.\ \eqref{Gdef} gives the conductance
\begin{equation}
G_{12}=\tfrac{1}{2}{\rm Tr}\,{\cal T}_{1}S_{dw}^{T}{\cal T}_{2}S_{dw},\label{Gdef2}
\end{equation}
in terms of the $2\times 2$ matrices ${\cal T}_{1}=t_{1}\Sigma_{y}t_{1}^{T}$, ${\cal T}_{2}={t'}_{2}^{T}\Sigma_{y}t'_{2}$. We now use the identity
\begin{equation}
{\rm Tr}\,A_{1}A_{2}=\tfrac{1}{2}\left({\rm Tr}\,A_{1}\sigma_{y}\right)\left({\rm Tr}\,A_{2}\sigma_{y}\right),\label{identity}
\end{equation}
valid for any pair of $2\times 2$ antisymmetric matrices $A_{1},A_{2}$. Taking $A_{1}={\cal T}_{1}$, $A_{2}=S_{dw}^{T}{\cal T}_{2}S_{dw}$ we arrive at
\begin{subequations}\label{RGresult}
\begin{align}
&G_{12}=(-1)^{p}\alpha_{1}\alpha_{2},\;\;\alpha_{i}=\tfrac{1}{2}{\rm Tr}\,{\cal T}_{i}\sigma_{y},\label{Gg1g2}\\
&R_{12}=(-1)^{p}\beta_{1}\beta_{2},\;\;\beta_{i}=\alpha_{i}/G_{i},\label{Rg1g2}
\end{align}
\end{subequations}
since ${\rm Tr}\,S_{dw}^{T}{\cal T}_{2}S_{dw}\sigma_{y}=(-1)^{p}{\rm Tr}\,{\cal T}_{2}\sigma_{y}$ in view of Eq.\ \eqref{Salpha}.

Eq.\ \eqref{RGresult} expresses the nonlocal conductance and resistance in terms of the scattering matrices $S_{1},S_{2}$ of the two contacts $N_{1},N_{2}$. The scattering matrix $S_{dw}$ of the domain wall enters only through the parity index $p$, and not through the mixing angle $\psi$. That the transferred charge depends only on a parity index is a generic feature of a single-mode scattering problem with class D symmetry \cite{Ber09a,Fu09,Akh09,Law09,Ber09b}. Quite generally, $p$ counts the number (modulo $2$) of zero-energy bound states, which in our case would be trapped in vortices in the domain wall.

A measurement of the domain wall conductance would have several characteristic features: Most prominently, the conductance is zero unless both contacts $N_{1}$ and $N_{2}$ are at the domain wall; if at least one contact is moved away from the domain wall, the conductance vanishes because a single Majorana edge mode cannot carry an electrical current at the Fermi level \cite{note2}. This feature would distinguish chiral \textit{p}-wave superconductors (symmetry class D) from chiral \textit{d}-wave superconductors (symmetry class C), where the Majorana edge modes come in pairs and can carry a current. The chirality itself can be detected by interchanging the injecting and detecting contacts: only one choice can give a nonzero conductance. While vortices trapped in the domain wall can change the sign of the conductance (through the parity index $p$), other properties of the domain wall have no effect on $G_{12}$. In particular, there is no dependence on the length $W$. 

To illustrate these features in a model calculation, we consider the case of two single-mode contacts ($N=N'=1$) coupled to the domain wall through a disordered interface. We model the effect of disorder using random contact scattering matrices $S_{1}$ and $S_{2}$, drawn independently with a uniform distribution from the ensemble of $4\times 4$ orthogonal matrices. In the context of random-matrix theory \cite{Meh04}, uniformly distributed ensembles of unitary matrices are called ``circular'', so our ensemble could be called the ``circular real ensemble'' (CRE) --- to distinguish it from the usual circular unitary ensemble (CUE) of complex unitary matrices \cite{note1}.

\begin{figure}[tb]
\centerline{\includegraphics[width=0.8\linewidth]{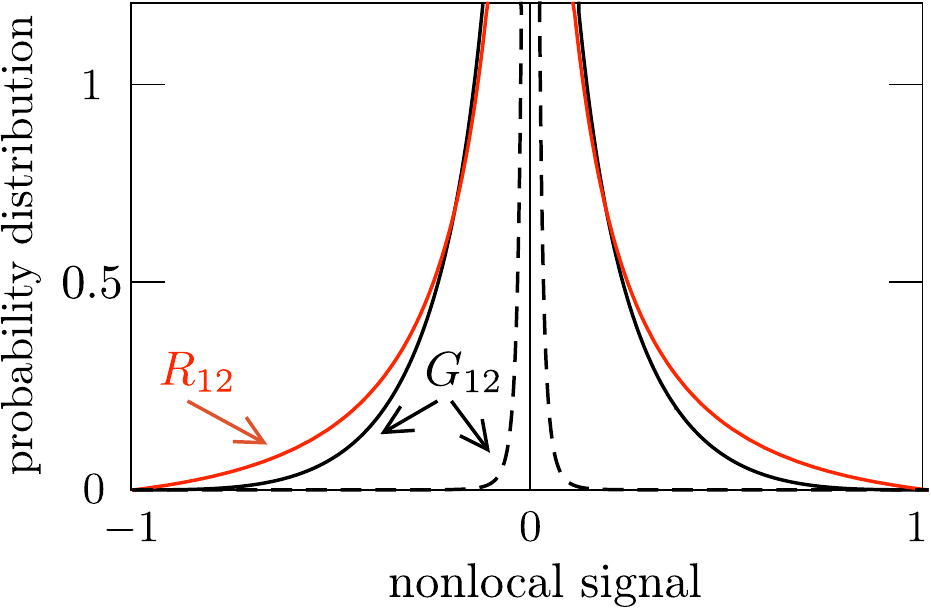}}
\caption{\label{fig_pg}
Solid curves: probability distributions of the nonlocal conductance $G_{12}$ (in units of $e^{2}/h$) and nonlocal resistance $R_{12}$ (in units of $h/e^{2}$). These are results for a random distribution of the $4\times 4$ orthogonal scattering matrices $S_{1}$ and $S_{2}$. The dashed curve shows the narrowing effect on $P(G_{12})$ of a tunnel barrier in both contacts (tunnel probability $\tau=0.1$). In contrast, $P(R_{12})$ is not affected by a tunnel barrier. 
}
\end{figure}

Using the expression for the uniform measure on the orthogonal group \cite{Ber09b,app}, we obtain the distributions of the parameters $\alpha_{i}$ and $\beta_{i}$ characterizing contact $N_{i}$:
\begin{equation}
P(\alpha)=1-|\alpha|,\;\;P(\beta)=(1+|\beta|)^{-2},\;\;|\alpha|,|\beta|\leq 1.\label{Palphabeta}
\end{equation}
The distribution of the nonlocal conductance $G_{12}=(-1)^{p}\alpha_{1}\alpha_{2}$, plotted in Fig.\ \ref{fig_pg}, then follows from
\begin{align}
P(G_{12})&=\int_{-1}^{1}d\alpha_{1}\int_{-1}^{1}d\alpha_{2}\,\delta(G_{12}-\alpha_{1}\alpha_{2})P(\alpha_{1})P(\alpha_{2})\nonumber\\
&=4|G_{12}|-4-2(1+|G_{12}|)\ln|G_{12}|,\;\;|G_{12}|<1.\label{pgresult}
\end{align}
(There is no dependence on the parity index $p$ because $P$ is symmetric around zero.) The distribution of the nonlocal resistance $R_{12}=(-1)^{p}\beta_{1}\beta_{2}$ follows similarly and as we can see in Fig.\ \ref{fig_pg} it lies close to $P(G_{12})$.

The difference between the two quantities $G_{12}$ and $R_{12}$ becomes important if the contacts between the metal and the superconductor contain a tunnel barrier. A tunnel barrier suppresses $G_{12}$ but has no effect on $R_{12}$. More precisely \cite{app}, any series resistance in the single-mode contacts $N_{1}$ and $N_{2}$ which does not couple electrons and holes drops out of the nonlocal resistance $R_{12}$. This remarkable fact is again a consequence of the product rule \eqref{identity}, which allows to factor a series conductance into a product of conductances. A tunnel barrier in contact $i$ then appears as a multiplicative factor in $\alpha_{i}$ and $G_{i}$, and thus drops out of the ratio $\beta_{i}=\alpha_{i}/G_{i}$ determining $R_{12}$.

To demonstrate the effect of a tunnel barrier (tunnel probability $\tau$), we have calculated the distribution of $\alpha$ using the Poisson kernel of the CRE \cite{Ber09c}, with the result
\begin{equation}
P(\alpha,\tau)=\frac{\tau^{2}}{[\tau+(1-\tau)|\alpha|]^{3}}-\frac{\tau^{2}|\alpha|}{[\tau+(1-\tau)\alpha^{2}]^{2}}.\label{Palphatau}
\end{equation}
The distribution of $\beta$ remains given by Eq.\ \eqref{Palphabeta}, independent of $\tau$. The dashed curves in Fig.\ \ref{fig_pg} show how the resulting distribution of the nonlocal conductance becomes narrowly peaked around zero for small $\tau$, in contrast to the distribution of the nonlocal resistance.

Among the various candidate systems for chiral \textit{p}-wave superconductivity, the recent proposal \cite{Sau09} based on the proximity effect in a semiconducting two-dimensional electron gas seems particularly promising for our purpose. Split-gate quantum point contacts (fabricated with well-established technology) could serve as single-mode injector and detector of electrical current. The chirality of the superconducting domains is determined by the polarity of an insulating magnetic substrate, so the location of the domain wall could be manipulated magnetically. The appearance of a nonlocal signal between the two point contacts would detect the domain wall and the disappearance upon interchange of injector and detector would demonstrate the chirality. 

As a direction for further research, we note that domains of opposite chirality are formed spontaneously in disordered samples. Since, as we have shown here, domain walls may carry electric current, a network of domain walls contributes to the conductivity and may well play a role in the anomalous (parity violating) current-voltage characteristic reported recently \cite{Nob09}.

We thank J. Nilsson for discussions. This research was supported by the Dutch Science Foundation NWO/FOM and by an ERC Advanced Investigator Grant.

\newpage

\appendix

\section{Averages over the circular real ensemble}
\label{nonlocal}

To calculate the distributions \eqref{Palphabeta} of the parameters $\alpha_{i}$ and $\beta_{i}$ we need the probability distribution of the $4\times 4$ scattering matrix $S_{i}$ of contact $i=1,2$ in the CRE. We may either work in the basis of electron and hole states, as in Ref.\ \cite{Ber09b}, or in the basis of Majorana states. Here we give a derivation of Eq.\ \eqref{Palphabeta} using the latter basis (which is the basis we used in the main text).

A $4\times 4$ orthogonal scattering matrix has the polar decomposition
\begin{align}
&S=
\begin{pmatrix}
e^{i\phi_{1}\sigma_{y}}&0\\
0&e^{i\phi_{2}\sigma_{y}}
\end{pmatrix}
\begin{pmatrix}
{\cal S}&{\cal C}\\
(-1)^{p+1}{\cal C}&(-1)^{p}{\cal S}
\end{pmatrix}\nonumber\\
&\qquad\cdot\begin{pmatrix}
e^{i\phi_{3}\sigma_{y}}&0\\
0&e^{i\phi_{4}\sigma_{y}}
\end{pmatrix},\label{polara}\\
&{\cal C}=
\begin{pmatrix}
\cos\psi_{1}&0\\
0&\cos\psi_{2}
\end{pmatrix},\;\;
{\cal S}=
\begin{pmatrix}
\sin\psi_{1}&0\\
0&\sin\psi_{2}
\end{pmatrix},
\label{polarb}
\end{align}
in terms of six real angles. We need the uniform measure on the orthogonal group, which defines the probability distribution in the circular real ensemble (CRE). This calculation proceeds along the same lines as in Ref.\ \cite{Ber09b} (where a different parametrization, in the electron-hole basis, was used). The result is that the angles $\phi_{1},\phi_{2},\phi_{3},\phi_{4}$ are uniformly distributed in $(0,2\pi)$, while the angles $\psi_{1},\psi_{2}$ have the distribution
\begin{equation}
P(\psi_{1},\psi_{2})=\tfrac{1}{4} |\cos^{2} \psi _{1}-\cos^{2} \psi _{2}|,\;\;0<\psi_{1},\psi_{2}<\pi.\label{psidist}
\end{equation}

We can now obtain the joint distribution $P(\alpha_{i},G_{i})$ of the injection (or detection) efficiency $\alpha_{i}$ and the (dimensionless) contact conductance $G_{i}$ of contact $i$. (We drop the label $i$ for ease of notation.) By definition,
\begin{align}
&\alpha=\tfrac{1}{2}{\rm Tr}\,t\sigma_{y}t^{T}\sigma_{y}=\cos\psi_{1}\cos\psi_{2},\label{alpha2def}\\
&G=1-\tfrac{1}{2}{\rm Tr}\,r\sigma_{y}r^{T}\sigma_{y}=1-\sin\psi_{1}\sin\psi_{2}.\label{G2def}
\end{align}
Notice the trigonometric inequality
\begin{equation}
0\leq|\alpha|\leq G\leq 2-|\alpha|.\label{inequality}
\end{equation}
By averaging over the CRE we find, remarkably enough, that the joint distribution of $\alpha$ and $G$ is uniform when constrained by this inequality,
\begin{align}
P(\alpha,G)&=\int_{0}^{\pi}d\psi_{1}\int_{0}^{\pi}d\psi_{2}\,P(\psi_{1},\psi_{2})\nonumber\\
&\times\delta(\alpha-\cos\psi_{1}\cos\psi_{2})\delta(G-1+\sin\psi_{1}\sin\psi_{2})\nonumber\\
&=\left\{\begin{array}{ll}
1/2&{\rm if}\;\;0\leq|\alpha|\leq G\leq 2-|\alpha|,\\
0&{\rm elsewise}.
\end{array}\right.\label{PGalpharesult}
\end{align}

\begin{figure}[tb]
\centerline{\includegraphics[width=0.9\linewidth]{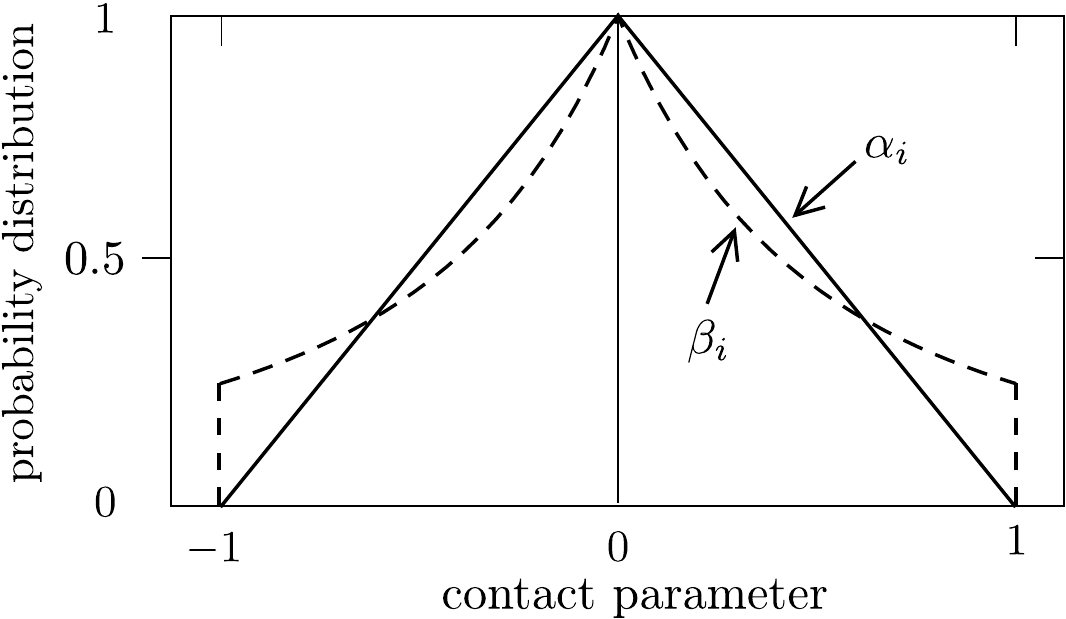}}
\caption{\label{fig_pab}
Probability distributions of the parameters $\alpha_{i}$ and $\beta_{i}=\alpha_{i}/G_{i}$ that characterize a single-mode contact in the CRE, given by Eqs.\ \eqref{Palpha2} and \eqref{Pbeta2}. The distribution \eqref{PG2} of $G_{i}-1$ is the same as that of $\alpha_{i}$, but these two quantities are not independent because of the inequality \eqref{inequality}.
}
\end{figure}

The marginal distributions of $\alpha$, $G$, and $\beta=\alpha/G$ now follow by integration over $P(\alpha,G)$,
\begin{align}
&P(\alpha)=1-|\alpha|,\;\;|\alpha|<1,\label{Palpha2}\\
&P_{G}(G)=1-|G-1|,\;\;0<G<2,\label{PG2}\\
&P(\beta)=(1+|\beta|)^{-2},\;\;|\beta|<1,\label{Pbeta2}
\end{align}
in accord with Eq.\ \eqref{Palphabeta}. We have plotted these distributions in Fig.\ \ref{fig_pab}.

\section{Proof that the tunnel resistance drops out of the nonlocal resistance}
\label{proof}

According to Eq.\ \eqref{RGresult}, the nonlocal conductance $G_{12}$ is determined by the product of the injection efficiency $\alpha_{1}$ of contact $N_{1}$ and the detection efficiency $\alpha_{2}$ of contact $N_{2}$. A tunnel barrier between the metal electrode and the superconductor suppresses the injection/detection efficiencies and thereby suppresses the nonlocal conductance.

The nonlocal resistance $R_{12}$ is determined by the ratio $\alpha_{i}/G_{i}$ of the injection/detection efficiency and the contact conductance $G_{i}$. Since both $\alpha_{i}$ and $G_{i}$ are suppressed by a tunnel barrier, one might hope that $R_{12}$ would remain of order $e^{2}/h$. In this Appendix we investigate the effect of a tunnel barrier on the nonlocal resistance, and demonstrate that it drops out identically for a single-mode contact between the normal metal and the superconductor.

The key identity that we will use to prove this cancellation, is the product rule \eqref{identity} and two corollaries:
\begin{subequations}\label{productrule}
\begin{align}
&\tfrac{1}{2}{\rm Tr}\,\bigl(\prod_{i}M_{i}\bigr)\sigma_{y}\bigl(\prod_{i}M_{i}\bigr)^{T}\sigma_{y}=\prod_{i}\bigl(\tfrac{1}{2}{\rm Tr}\,M_{i}\sigma_{y}M_{i}^{T}\sigma_{y}\bigr),\label{productrule1}\\
&\tfrac{1}{2}{\rm Tr}\,(M\sigma_{y}M^{T}\sigma_{y})^{-1}=\bigl[\tfrac{1}{2}{\rm Tr}\,M\sigma_{y}M^{T}\sigma_{y}\bigr]^{-1},\label{productrule2}
\end{align}
\end{subequations}
valid for arbitrary $2\times 2$ matrices $M_{i}$.

Considering any one of the two contacts, we assume that its scattering matrix $S_{0}$ is modified by a tunnel barrier with scattering matrix $\delta S$. Transmission and reflection submatrices are defined as in Eq.\ \eqref{S1def},
\begin{equation}
S_{0}=
\begin{pmatrix}
r_{0}&t_{0}\\
t'_{0}&r'_{0}
\end{pmatrix},\;\;
\delta S=
\begin{pmatrix}
\delta r&\delta t\\
\delta t'&\delta r'
\end{pmatrix}.
\label{S0deltaSdef}
 \end{equation}
For a single-mode contact, each submatrix has dimension $2\times 2$. Both $S_{0}$ and $\delta S$ are real orthogonal matrices at zero energy (in the basis of Majorana fermions). The tunnel barrier does not couple electrons and holes, which means that the submatrices of $\delta S$ must commute with $\sigma_{y}$,
\begin{equation}
[\sigma_{y},\delta r]=[\sigma_{y},\delta r']=[\sigma_{y},\delta t]=[\sigma_{y},\delta t']=0.\label{commute}
\end{equation}
The submatrices of $S_{0}$ are not so constrained.

The total scattering matrix $S$ of the contact is constructed from $S_{0}$ and $\delta S$, according to the composition rule for scattering matrices. The transmission and reflection submatrices of $S$ take the form
\begin{subequations}\label{rtrt}
\begin{align}
&t=t_{0}(1-\delta rr'_{0})^{-1}\delta t,\label{rtrta}\\
&t'=\delta t'(1-r'_{0}\delta r)^{-1}t'_{0},\label{rtrtb}\\
&r'=\delta r'+\delta t'r'_{0}(1-\delta rr'_{0})^{-1}\delta t,\label{rtrtc}\\
&r=r_{0}+t_{0}\delta r(1-r'_{0}\delta r)^{-1}t'_{0}\label{rtrtd}.
\end{align}
\end{subequations}

The injection efficiency $\alpha$ and detection efficiency $\alpha'$ are defined by
\begin{equation}
\alpha=\tfrac{1}{2}{\rm Tr}\,t\sigma_{y}t^{T}\sigma_{y},\;\;
\alpha'=\tfrac{1}{2}{\rm Tr}\,t'\sigma_{y}{t'}^{T}\sigma_{y}.
\end{equation}
Using the identities \eqref{productrule1} and \eqref{productrule2} we can factor these quantities,
\begin{equation}
\alpha=\alpha_{0}\delta\alpha/X,\;\;
\alpha'=\alpha'_{0}\delta\alpha'/X,\label{alphafactor}
\end{equation}
into the product of the injection/detection efficiencies $\alpha_{0},\alpha'_{0}$ without the tunnel barrier and terms containing the effect of the tunnel barrier:
\begin{subequations}\label{Xdef}
\begin{align}
&\alpha_{0}=\tfrac{1}{2}{\rm Tr}\,t_{0}\sigma_{y}t_{0}^{T}\sigma_{y},\;\;
\alpha'_{0}=\tfrac{1}{2}{\rm Tr}\,t'_{0}\sigma_{y}{t'}_{0}^{T}\sigma_{y},\\
& \delta\alpha=\tfrac{1}{2}{\rm Tr}\,\delta t\sigma_{y}\delta t^{T}\sigma_{y},\;\;
\delta\alpha'=\tfrac{1}{2}{\rm Tr}\,\delta t'\sigma_{y}\delta {t'}^{T}\sigma_{y},\\
&X=\tfrac{1}{2}{\rm Tr}\,(1-\delta rr'_{0})\sigma_{y}(1-\delta rr'_{0})^{T}\sigma_{y}.
\end{align}
\end{subequations}

Since $\delta t$ and $\delta t'$ commute with $\sigma_{y}$, the terms $\delta\alpha$, $\delta\alpha'$ simplify to
\begin{equation}
\delta\alpha=\delta\alpha'=\tfrac{1}{2}{\rm Tr}\,\delta t\delta t^{T},
\end{equation}
where we have used the orthogonality condition, $\delta S^{T}\delta S=\delta S\delta S^{T}=1$, to equate the traces of $\delta t\delta t^{T}$ and $\delta t'\delta {t'}^{T}$. The term $X$ can similarly be reduced to
\begin{equation}
X=1+(1-\delta\alpha)(1-G_{0})-{\rm Tr}\,\delta rr'_{0},\label{Xsimple}
\end{equation}
where $G_{0}$ is the contact conductance (in units of $e^{2}/h$) in the absence of the tunnel barrier:
\begin{equation}
G_{0}=\tfrac{1}{2}{\rm Tr}\,(1-r'_{0}\sigma_{y}{r'}_{0}^{T}\sigma_{y}).\label{G0def}
\end{equation}

We now turn to the contact conductances $G_{i}$, in order to show that the effect of the tunnel barrier is contained in the same factor $\delta\alpha/X$ (which will then cancel out of the ratio $\beta_{i}=\alpha_{i}/G_{i}$). Considering again a single contact, and dropping the index $i$ for ease of notation, we start from the definition of the contact conductance (in units of $e^{2}/h$):
\begin{equation}
G=\tfrac{1}{2}{\rm Tr}\,(1-r'\sigma_{y}{r'}^{T}\sigma_{y}).\label{Gdefapp}
\end{equation}
We substitute Eq.\ \eqref{rtrtc}, and try to factor out the terms containing the transmission and reflection matrices of the tunnel barrier.

It is helpful to first combine the two terms in Eq.\ \eqref{rtrtc} into a single term, using the orthogonality of $\delta S$:
\begin{align}
r'&=-({\delta t'}^{T})^{-1}\delta r^{T}\delta t+\delta t'r'_{0}(1-\delta rr'_{0})^{-1}\delta t\nonumber\\
&=({\delta t'}^{T})^{-1}(r'_{0}-\delta r^{T})(1-\delta rr'_{0})^{-1}\delta t.\label{rprime}
\end{align}
We now substitute Eq.\ \eqref{rprime} into Eq.\ \eqref{Gdef} and use the identities \eqref{productrule} to factor the trace,
\begin{align}
G&=1-X^{-1}\tfrac{1}{2}{\rm Tr}\,(r'_{0}-\delta r^{T})\sigma_{y}({r'}_{0}^{T}-\delta r)\sigma_{y}\nonumber\\
&=1-X^{-1}(2-\delta\alpha-G_{0}-{\rm Tr}\,\delta rr'_{0}),\label{Gsimple}
\end{align}
where we also used the commutation relations \eqref{commute}. The remaining trace of $\delta rr'_{0}$ can be eliminated with the help of Eq.\ \eqref{Xsimple}, and so we finally arrive at the desired result:
\begin{equation}
G=G_{0}\delta\alpha/X.\label{G0factor}
\end{equation}


\begin{thebibliography}{99}
\bibitem{note2} That the nonlocal conductance vanishes if one of the two contacts couples only to a single domain, can be seen directly from Eq.\ \eqref{RGresult}: If, say, contact 1 couples only to the right domain, then only the $2,2$ element of ${\cal T}_{1}$ can be nonzero, but since this matrix is antisymmetric the $2,2$ element must also vanish and ${\cal T}_{1}$ must be zero identically. This implies $\alpha_{1}=0$, hence $G_{12}=0$.
\bibitem{note1} The name ``circular orthogonal ensemble'' (COE) might be more appropriate for the ensemble of uniformly distributed orthogonal matrices, but this name is already in use for the ensemble of unitary symmetric matrices.
\bibitem{app} For a detailed calculation we refer to the appendices.
\end{thebibliography}
\end{document}